\documentclass[
    reprint,
    pra,
    aps,
    superscriptaddress,
]{revtex4-2}

\usepackage{xspace}
\usepackage{graphicx} %
\graphicspath{{./fig}}

\usepackage[dvipsnames,table,xcdraw]{xcolor}
\definecolor{rmpblue}{HTML}{2e3092}
\usepackage[
	colorlinks=true,
	citecolor=rmpblue,
	linkcolor=rmpblue,
	filecolor=magenta,
	urlcolor=rmpblue,
]{hyperref}

\usepackage{physics}
\usepackage{siunitx}

\newcommand{\iu}{\mathrm{i}\mkern1mu}

\newcommand{\affilANU}{Nonlinear Physics Center, Research School of Physics, Australian National University, Canberra ACT 2601, Australia}

\newcommand{\affilSUSTech}{Department of Materials Science and Engineering \& Institute for Applied Optics and Precision Engineering, Southern University of Science and Technology, Shenzhen, 518055 P. R. China}

\newcommand{\affilRIKEN}{Metaphotonics Research Team, RIKEN Center for Advanced Photonics, Saitama 351-0198, Japan}

\newcommand{\affilSeoul}{Department of Physics and Astronomy and Institute of Applied Physics, Seoul National University, Seoul 08826, Republic of Korea}

\begin{document}

\author{Pavel Tonkaev}
\altaffiliation{Contributed equally}
\affiliation{\affilANU}
\email{pavel.tonkaev@anu.edu.au}

\author{Yeqi Zhuang}
\altaffiliation{Contributed equally}
\affiliation{\affilSUSTech}

\author{Donghwee Kim}
\affiliation{\affilSeoul}

\author{Ivan Toftul}
\affiliation{\affilANU}

\author{Takeshi Yamaguchi}
\affiliation{\affilRIKEN}

\author{Kingfai Li}
\affiliation{\affilSUSTech}

\author{Jiaming Huang}
\affiliation{\affilSUSTech}

\author{Heng Wang} 
\affiliation{\affilSUSTech}

\author{Takuo Tanaka}
\affiliation{\affilRIKEN}

\author{Hong-Gyu Park}
\affiliation{\affilSeoul}

\author{Guixin Li}
\affiliation{\affilSUSTech}
\email{ligx@sustech.edu.cn}

\author{Yuri Kivshar}
\affiliation{\affilANU}
\email{yuri.kivshar@anu.edu.au}

\title{Nonlinear chiral response from linearly achiral membrane metasurfaces}

\begin{abstract}
Chiral photonics aims to control and engineer light handedness for many applications in optical communications, biological and chemical sensing, and quantum technologies. While traditional approaches focus on engineering strong linear chiroptical response, nonlinear chiral phenomena remain largely unexplored. Here, we demonstrate experimentally a pronounced {\it nonlinear chiral response} in free-standing silicon membrane metasurfaces that are effectively achiral in the linear regime. By employing patterned membranes with both $C_4$-symmetry and intentionally broken in-plane symmetry, we reveal that strong nonlinear circular dichroism can be observed in third-harmonic generation. An unperturbed metasurface exhibits strong cross-polarized third-harmonic signal with nonlinear circular dichroism of the value -0.83, whereas in-plane symmetry breaking enables a co-polarized channel, and it reverses the sign of nonlinear circular dichroism that may be as large as the value 0.41. Our findings suggest a novel approach for engineering nonlinear chiral responses in metasurfaces, complementing traditional approaches and paving the way towards advanced chiral metadevices. 
\end{abstract}

\maketitle

\section{Introduction}

\begin{figure}
    \centering
    \includegraphics[width=1\linewidth]{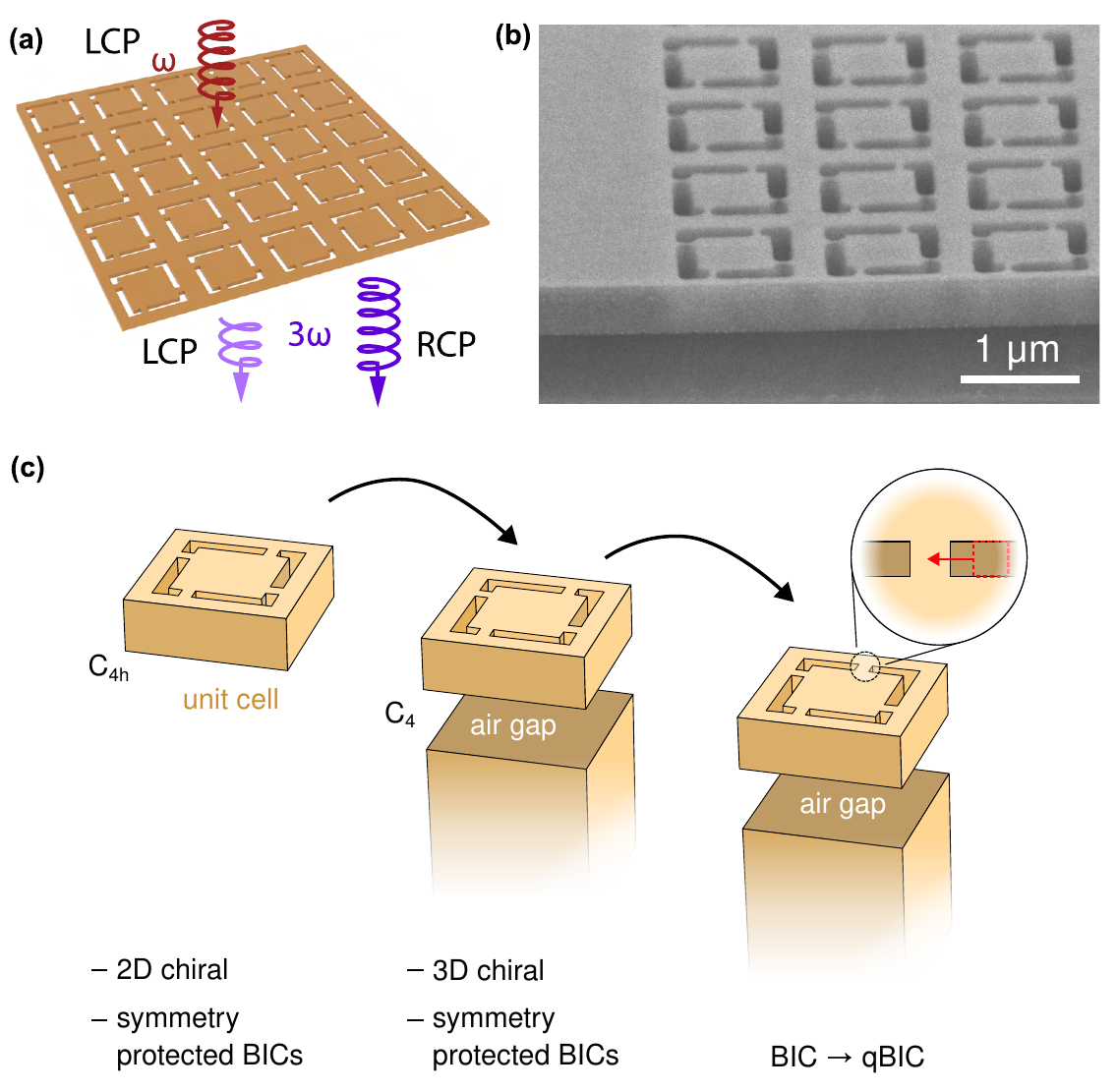}
    \caption{ \textbf{Nonlinearly chiral metasurfaces}. (a) Schematic illustration of a membrane metasurface designed for chiral third-harmonic generation. (b) Scanning electron microscope (SEM) image of the fabricated metasurface. 
    (c) Degradation of properties of a membrane in free space when transitioning to a less symmetric cases. 2D chiral ($C_{4h}$ symmetric): $T_{\text{RR}} = T_{\text{LL}}$, $T_{\text{RL}} = T_{\text{LR}} = 0$, and 
    $I_{\text{RR}}^{(3\omega)} = I_{\text{LL}}^{(3\omega)} = 0$.
    3D chiral ($C_4$ symmetric): same as 2D chiral but linear co-polarized CD is not forbidden.
    3D chiral (asymmetric): no symmetry protected BIC.~\cite{koshelev2024scattering}.
    }
    \label{fig1}
\end{figure}

Chiral photonics harnesses the selective interaction of circularly polarized light with structured matter and underpins a wide range of emerging technologies. This includes secure optical communications~\cite{zhu2025polarization}, subwavelength imaging~\cite{kfir2017nanoscale}, biosensing~\cite{hendry2010ultrasensitive} and even brain wave detection to enhance human effectiveness. Central to these applications is the ability to achieve and control optical activity manifested in circular dichroism (CD) in visible and near-infrared spectral ranges. However, the chiroptical response of naturally occurring materials is extremely weak, typically several orders of magnitude smaller than the requirement for practical use in compact photonics systems.

To overcome this limit, artificial chiral materials with tailored optical properties have been developed. Subwavelength-structured materials have emerged as an important tool to achieve macroscopic chirality and enhance weak natural chiroptical effects by several orders of magnitude~\cite{pendry2004a,wang2009chiral,oh2015chiral}. Early designs focused on metallic plasmonic structures to achieve strong resonance-enhanced chiral response~\cite {valev2013chirality,hentschel2017chiral}. Despite their initial success, plasmonic systems suffer from intrinsic Ohmic losses, motivating a transition toward the study of dielectric structures, in particular those composed of ordered arrays of subwavelength resonators\cite{gorkunov2020metasurfaces,overvig2021chiral,kuhner2023unlocking}. These dielectric platforms offer low-loss operation and access to high-$Q$ resonances, enabling a new class of compact chiral photonic devices. The enhanced control of light-matter interactions in dielectric metasurfaces has enabled a range of chiral functionalities, including chiral emission~\cite{zhang2022chiral,gromyko2024unidirectional}, chiral sensing~\cite{mohammadi2018nanophotonic, solomon2018enantiospecific}, and chiral harmonic generation~\cite{koshelev2023nonlinear}. These developments have largely focused on maximising \textit{linear chiroptical response} of nanostructures. 

Building on these efforts, \textit{nonlinear chiral metaphotonics} has recently emerged as a promising approach that leverages symmetry principles and resonances to control harmonics generation in chiral nanostructures~\cite{koshelev2023nonlinear}. This rapidly developing field lies at the intersection of nanophotonics, symmetry engineering, and nonlinear optics, aiming to control the handedness of light at the nanoscale. Recent advances have demonstrated both linear and nonlinear chiroptical effects enhanced by high-$Q$ resonances~\cite{gandolfi2021near,shi2022planar,koshelev2023resonant,tonkaev2024nonlinear} and governed by the intrinsic and extrinsic symmetry properties of the metasurfaces~\cite{chen2014symmetry,konishi2020circularly,koshelev2024scattering,toftul2024chiral,toftul2025monoclinic}. Typically, a strong nonlinear chiral response is achieved by designing metasurfaces that already exhibit chirality in the linear regime, generally achieved through breaking the inversion symmetry, leading to pronounced circular dichroism at both fundamental and harmonic frequencies.

In contrast, here we propose and experimentally realise a completely different strategy that enables purely nonlinear circular dichroism. In our concept, the photonic structure is designed to be effectively achiral for the scattering of light in the linear regime, exhibiting negligible linear CD. However, through symmetry-selective nonlinear interactions, the same structure generates a strong chiral response in the nonlinear regime. Such “nonlinear chirality”  represents a departure from conventional strategies by decoupling linear and nonlinear chiral behaviour, and it opens novel opportunities and unexpected applications, such as nonlinear imaging, encoding, and chiral-selective sensing. In such scenarios,   circular dichroism is exclusively detected in the harmonic fields, whereas the structure remains achiral in the linear regime.

\section{Realisation of nonlinear chiral metasurface}

{To realise circular dichroism solely in the nonlinear response while suppressing it completely in the linear transmission, we base our design on \textit{symmetry principles} and related selection rules. Nonlinear CD emerges in structures lacking in-plane inversion symmetry (i.e.,  2D chiral)~\cite{koshelev2024scattering}. At the same time, by preserving up–down mirror symmetry, we ensure identical co-polarized transmission for RCP and LCP waves in the linear regime. Finally, to eliminate any residual imbalance between the two polarizations, we impose $C_4$ rotational symmetry, which forbids cross-polarization~\cite{Kumar2025arXiv_chiral_membrane,koshelev2024scattering,gorkunov2020metasurfaces,sinev2025chirality}.
As the result of this reasoning, our metasurfaces are composed of free-standing silicon membranes patterned with square lattices of nanostructured unit cells, as illustrated in Figure~\ref{fig1}a. The metasurfaces were fabricated using electron beam lithography followed by a sequence of etching and release steps. Detailed information on the fabrication procedure is provided in Supporting Information. Each metasurface has lateral dimensions of 50~\textmu m$~\times~$50~\textmu m. The membranes are mechanically supported by SiO$_2$ anchors extending 50~\textmu m inward from all four edges. The vertical separation between the suspended membrane and the underlying silicon substrate is approximately 400~nm, enabling efficient optical isolation and interaction with the incident field. A side-view scanning electron microscope (SEM) image of the fabricated structure is shown in Figure~\ref{fig1}b. The metasurfaces are designed to support symmetry protected bound state in the continuum (BIC) in the infrared spectral region. In a fully symmetric, free-standing membrane, the out-of-plane mirror symmetry prohibits any chiral linear response~\cite{koshelev2024scattering}. However, in practical implementations, the presence of a supporting substrate beneath the air gap breaks this vertical symmetry, thereby lifting the constraint and enabling chiral response in the linear regime. Furthermore, to introduce a quasi-BIC resonance, we intentionally break the in-plane symmetry of the unit cell. Figure~\ref{fig1}c schematically summarises the key  structural features and the symmetry breaking steps. 

\begin{figure*}
    \centering
    \includegraphics[width=0.7\linewidth]{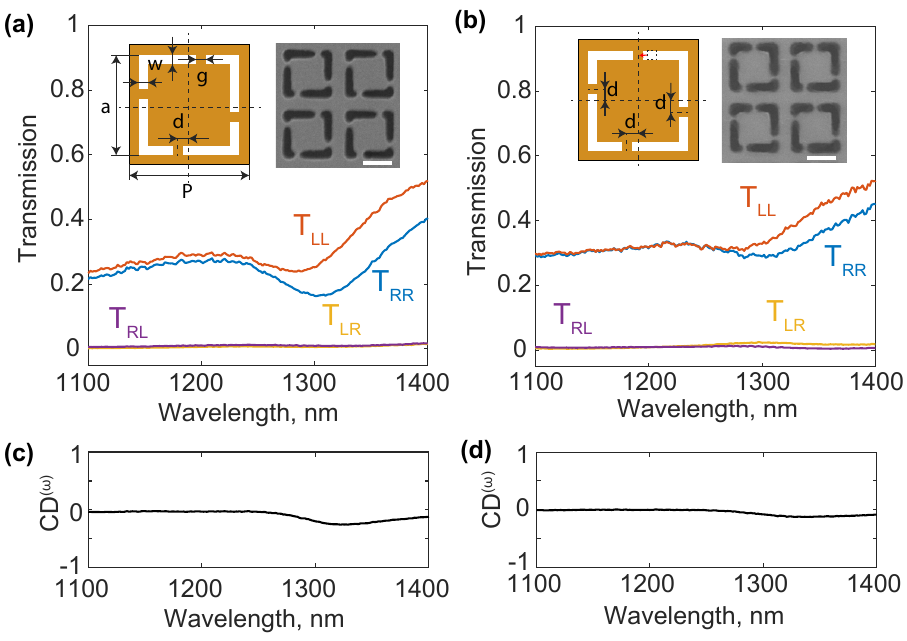}
    \caption{\textbf{Experimental Linear optical properties}. Transmission spectra of the unperturbed (a) and perturbed (b) metasurfaces under left- and right-circularly polarized light. Insets show schematic images of the corresponding unit cells with parameters and top-view SEM images. Linear circular dichroism spectra for the unperturbed (c) and perturbed (d) metasurfaces.}
    \label{fig2}
\end{figure*}

To explore the role of structural symmetry in optical behavior, we analyze two types of membrane metasurfaces: one possessing $C_4$ symmetry and another with intentionally broken $C_4$ symmetry. The electromagnetic response of a metasurface is defined by its eigenmodes, whose properties are determined by symmetry. In the $C_4$-symmetric case, certain modes  cannot couple to the far field, forming symmetry protected BICs~\cite{Kumar2025arXiv_chiral_membrane,Gladyshev2020PRB,Shalin2023}. Breaking the rotational symmetry, as illustrated in Fig.~\ref{fig1}(c), lifts this restriction and enables far-field coupling of these modes, thereby transforming them from BICs into quasi-BICs.
The corresponding structures are depicted in the insets of Figures~\ref{fig2}a and \ref{fig2}b, respectively. The metasurface unit cell is a square with a period $P$, consisting of L-shaped cavities of width $w$, separated by square gaps of size $g$ positioned to form a $C_4$-symmetric pattern. These gaps are placed at a distance $d$ from the central axis. To break the $C_4$ symmetry, one of the connecting bridges is shifted to the centerline, as shown in Figure~\ref{fig1}c. For the fabricated structures, the design parameters are: $P$ = 1000 nm, $a$ = 740 nm, $w$ = 80 nm, $g$ = 80 nm, and $d$ = 165 nm. The membrane thickness is 370 nm for both the cases. Top-view SEM images of the fabricated metasurfaces are also provided in  the insert of Figures~\ref{fig2}a and \ref{fig2}b.

We begin by investigating the linear optical properties of the metasurfaces through both numerical simulations and experiments. First, we numerically simulate the linear response by illuminating the metasurfaces with left-circularly polarized (LCP) and right-circularly polarized (RCP) plane waves. For each incident polarization, the transmitted LCP and RCP components are extracted. This allows us to determine the co-polarized transmission coefficients, $T_{\text{RR}}$ and $T_{\text{LL}}$, which represent the portions of light maintaining the LCP or RCP states, respectively. Additionally, the cross-polarized components, $T_{\text{RL}}$ and $T_{\text{LR}}$, corresponding to conversion between RCP and LCP, are also evaluated. The simulation details and obtained theoretical spectra are provided in Supporting Information. The simulated transmission spectra for unperturbed metasurface reveal that the co-polarized components $T_{\text{RR}}$ and $T_{\text{LL}}$ are equal, while cross-polarized components $T_{\text{RR}}$ and $T_{\text{LL}}$ vanish, as expected due to in-plane $C_4$ symmetry and out-of-plane mirror symmetry. In contrast, for the perturbed metasurface with broken in-plane symmetry, while the co-polarized components remain similar, the cross-polarized components become non-zero, indicating symmetry-induced polarization convention.

Experimentally, the metasurfaces are illuminated with LCP and RCP light, and the transmitted signals are analysed to extract both LCP and RCP components for each incident polarization. Experimental procedures are detailed in Supporting Information. Figure~\ref{fig2}a shows the transmission spectra for the metasurface with $C_4$ symmetry. The co-polarized transmission components under RCP ($T_{\text{RR}}$) and LCP illumination ($T_{\text{LL}}$) are shown by the blue and red curves, respectively, while the cross-polarized components, corresponding to converting RCP to LCP ($T_{\text{RL}}$) and LCP to RCP ($T_{\text{LR}}$), are represented by the purple and yellow curves, respectively. As expected, the cross-polarized transmission is negligible, whereas the co-polarized components are non-zero and closely match each other.

To understand the effect of the symmetry breaking, next we study experimentally linear optical properties of the metasurface with broken in-plane $C_4$ symmetry. The corresponding transmission spectra are presented in Figure~\ref{fig2}b. While the overall behaviour remains similar to the symmetric case, the cross-polarized transmission is slightly enhanced around 1300 nm. This increase is attributed to the in-plane symmetry breaking, which facilitates cross-polarization effects in transmission. According to the symmetry analyses and simulations, the co-polarized components should be identical for the two types of metasurfaces due to the out-of-plane symmetry. However, in the fabricated samples, the membranes are supported by a silicon substrate with an air gap of approximately 400 nm, which may perturb the symmetry, as discussed earlier. Additionally, the membrane walls in the actual structures are slightly inclined, further breaking the out-of-plane symmetry.

To characterize the difference in the metasurface response to RCP and LCP light, we use {\it linear circular dichroism} (CD), defined as:
\begin{equation}
\mathrm{CD}^{(\omega)} = \frac{(T_{\text{RR}} + T_{\text{RL}})  - (T_{\text{LL}}+T_{\text{LR}})}{T_{\text{RR}}+T_{\text{RL}} + T_{\text{LL}}+T_{\text{LR}}},    \label{eq1}
\end{equation}
where $T_{\alpha \beta}$ denotes the co-polarized and cross-polarized transmission coefficients. Here,  $\alpha$ refers to the input polarization (LCP (L) or RCP (R)), and $\beta$ refers to the detected polarization (LCP or RCP). The calculated CD spectra for the unperturbed and perturbed metasurfaces are shown in Figures~\ref{fig2}c and \ref{fig2}d, respectively. As seen, the CD remains close to zero in the short-wavelength region and becomes non-zero for wavelengths above 1300 nm. The maximum linear CD reaches a value of -0.25 for the unperturbed metasurface, while it reduces to -0.12 for the perturbed metasurface.

\begin{figure}
    \centering
    \includegraphics[width=0.75\linewidth]{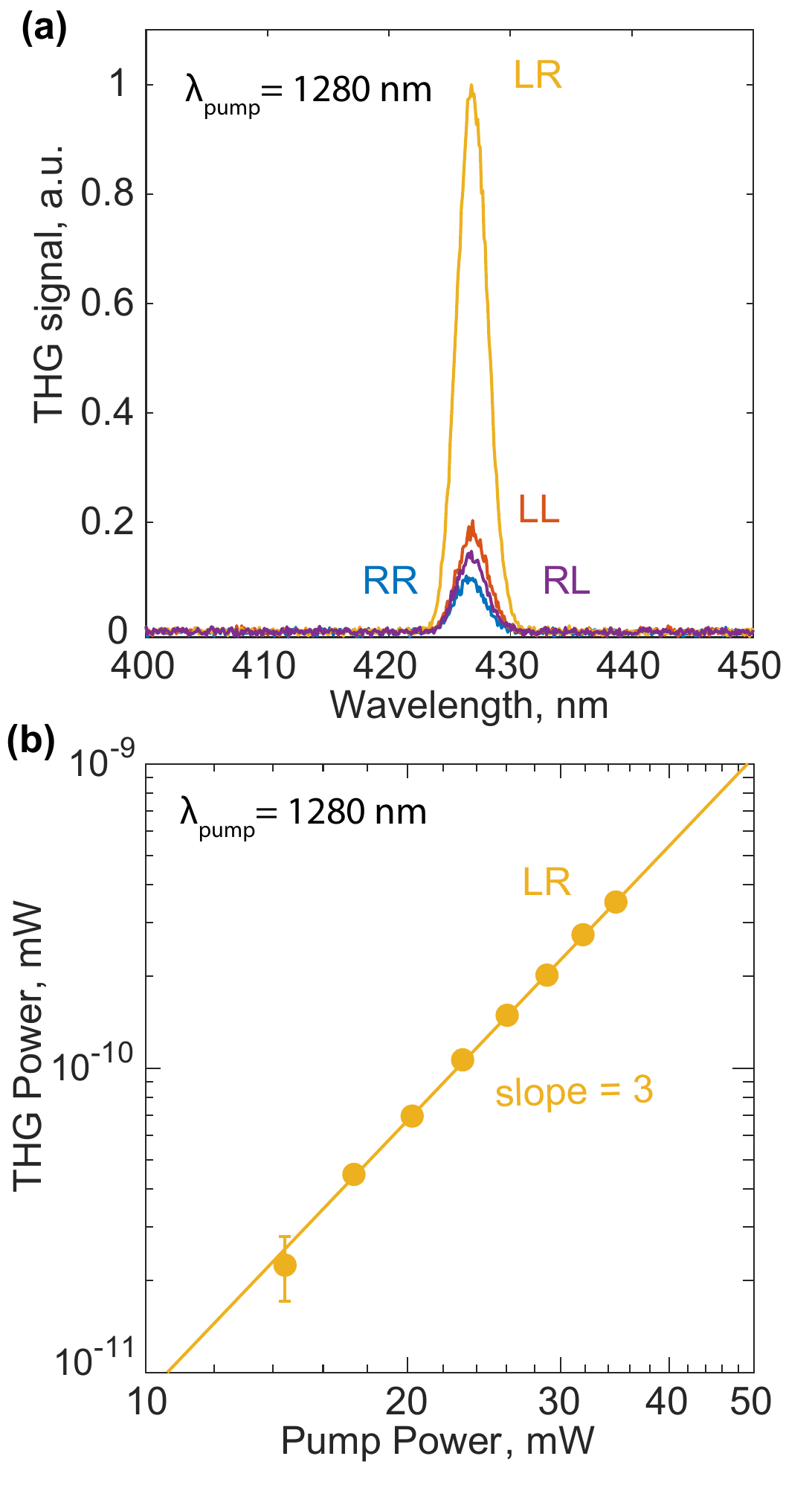}
    \caption{\textbf{Third-harmonic generation}. (a) Third-harmonic spectra from the unperturbed metasurface pumped at a wavelength of 1280 nm. Shown are co-polarized signals for RCP (blue) and LCP (red) excitation and detection, and cross-polarized signals for LCP (yellow) and RCP (purple) excitation. (b) Third-harmonic power as a function of pump power for the unperturbed metasurface under LCP excitation and RCP detection, with excitation wavelength at 1280 nm.}
    \label{fig3}
\end{figure}

We now turn to the nonlinear optical properties of the metasurfaces. They are pumped from the backside using a tunable femtosecond laser with LCP and RCP light, and the generated third-harmonic signal is analysed by separating its LCP and RCP components. Additional experimental details are provided in the Supporting Information. Figure~\ref{fig3}a presents the third-harmonic spectra from the unperturbed metasurface pumped at a wavelength of 1280 nm for different combinations of circular polarizations in the pump and collection paths. The dominant contribution is observed in the cross-polarized LR component, while the other components exhibit similar and significantly lower intensities. Figure~\ref{fig3}b shows the power dependence of the RCP component of the third-harmonic signal when the pump is LCP. The observed cubic dependence confirms the third-order nonlinear nature of the process. The maximum absolute third-harmonic generation efficiency achieved is 1.01$\cdot$10$^{-11}$, measured at a pump power of 34.7 mW.

\begin{figure}
    \centering
    \includegraphics[width=1\linewidth]{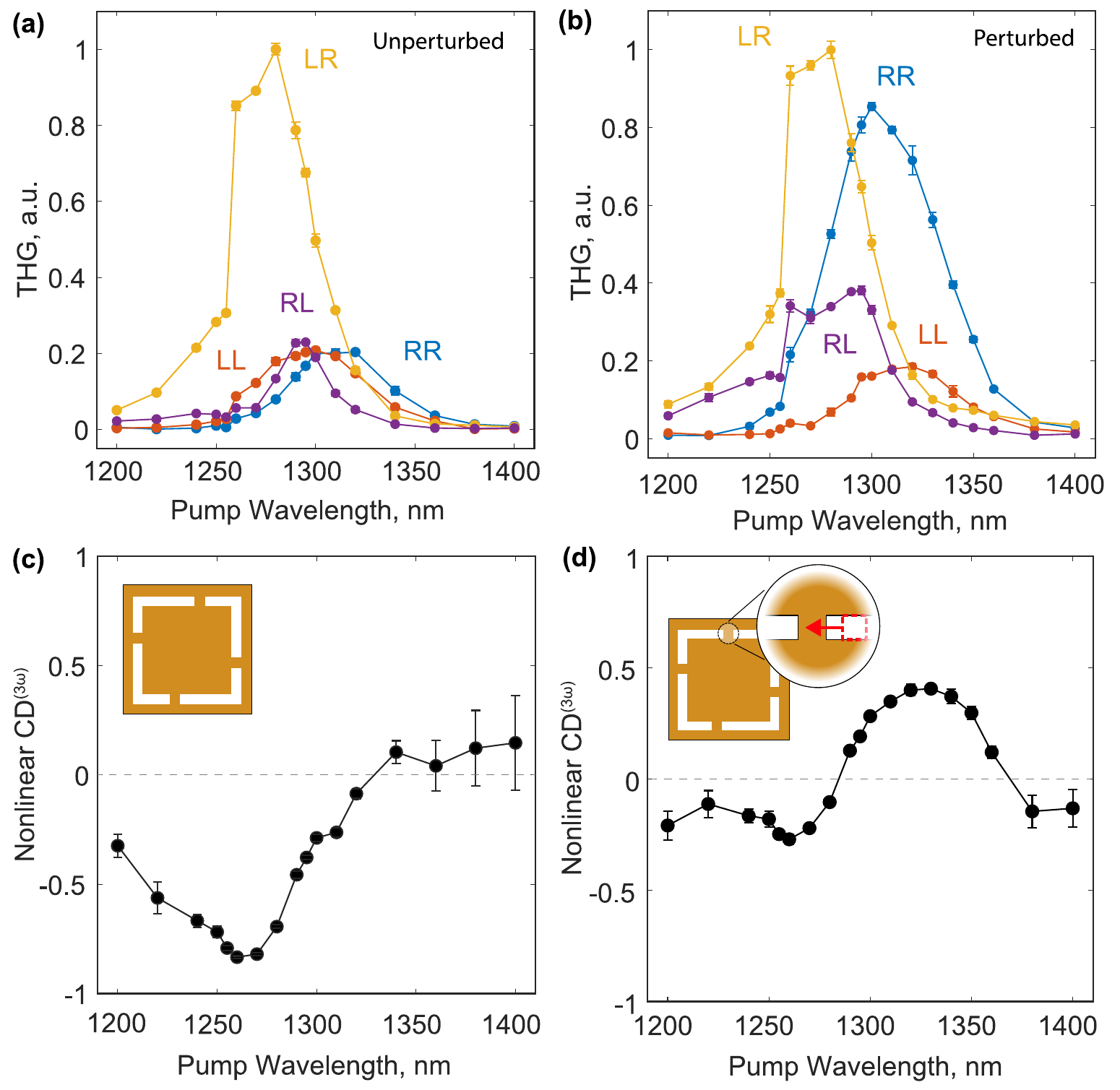}
    \caption{\textbf{Experimental chiral nonlinear response}. Third-harmonic signals for co-polarized excitation and detection with RCP (blue) and LCP (red), and for cross-polarized signals under LCP (yellow) and RCP (purple) excitation, measured from the unperturbed (a) and perturbed (b) metasurfaces. Third-harmonic circular dichroism (CD) spectra for the unperturbed (c) and perturbed (d) metasurfaces. Insets show schematic images of the corresponding unit cells.}
    \label{fig4}
\end{figure}

To explore spectral features of nonlinear chiral response, we further measure the THG from metasurfaces under excitation with different pump wavelengths. The metasurfaces are illuminated with a tunable femtosecond laser in the 1200 nm - 1400 nm range. For each pump wavelength, the corresponding THG signal is collected, and the maximum intensity is extracted. Figure~\ref{fig4}a presents the THG intensity measured for different circular polarizations of both the pump and the detected signal using the unperturbed metasurface. As shown, the THG is generally enhanced near 1300 nm for all polarization channels. The cross-polarized component LR reaches the maximum at 1280 nm wavelength and is approximately five times stronger than the co-polarized components and the cross-polarized component RL. This behavior is expected due to the $C_4$ symmetry of the metasurface, which permits only the cross-polarized THG~\cite{koshelev2024scattering}. To quantify the difference in nonlinear response between LCP and RCP pump, we define {\it nonlinear circular dichroism} $\mathrm{CD}^{(3\omega)}$ as: 
\begin{equation}
    \mathrm{CD}^{(3\omega)} = \frac{I_{\text{RR}}^{(3\omega)}+I_{\text{RL}}^{(3\omega)} - (I_{\text{LL}}^{(3\omega)}+I_{\text{LR}}^{(3\omega)})}{I_{\text{RR}}^{(3\omega)}+I_{\text{RL}}^{(3\omega)} + I_{\text{LL}}^{(3\omega)}+I_{\text{LR}}^{(3\omega)}},
    \label{eq2}
\end{equation}
where $I_{\alpha \beta}^{(3\omega)}$ denotes the THG intensity with $\alpha$  and $\beta$ indicating the circular polarization of the pump and the detected signal, respectively (LCP (L) or RCP (R)). The calculated nonlinear CD for the unperturbed metasurface is shown in Figure~\ref{fig4}c. The maximum CD value of -0.83 occurs at a pump wavelength of 1260 nm, corresponding to a spectral region where the LR component of the THG is strongly enhanced. Outside this region, the nonlinear CD approaches zero.

To understand how symmetry breaking modifies the nonlinear chiral response, we repeat the wavelength-dependent THG measurements from the perturbed metasurface for various combinations of circular polarizations in both the pump and the detected signal. The results are presented in Figure~\ref{fig4}b. As evident from the data, the cross-polarized LR component remains dominant at a pump wavelength of 1280 nm. However, as the wavelength increases to 1310 nm, the co-polarized component RR becomes significantly enhanced. This behaviour arises due to the intentional breaking of the in-plane $C_4$ symmetry, which enables coupling between the symmetry-forbidden nonlinear channels. To quantify the polarization-dependent nonlinear response, we calculate the nonlinear CD using Eq.~(\ref{eq2}). The resulting nonlinear CD spectrum is shown in Figure~\ref{fig4}d. Unlike the unperturbed metasurface, where a modest negative CD is observed near 1260 nm, the perturbed structure exhibits a pronounced change: the CD changes sign and reaches a positive peak value of 0.41 at 1330 nm.

\begin{figure}
    \centering
    \includegraphics[width=1\linewidth]{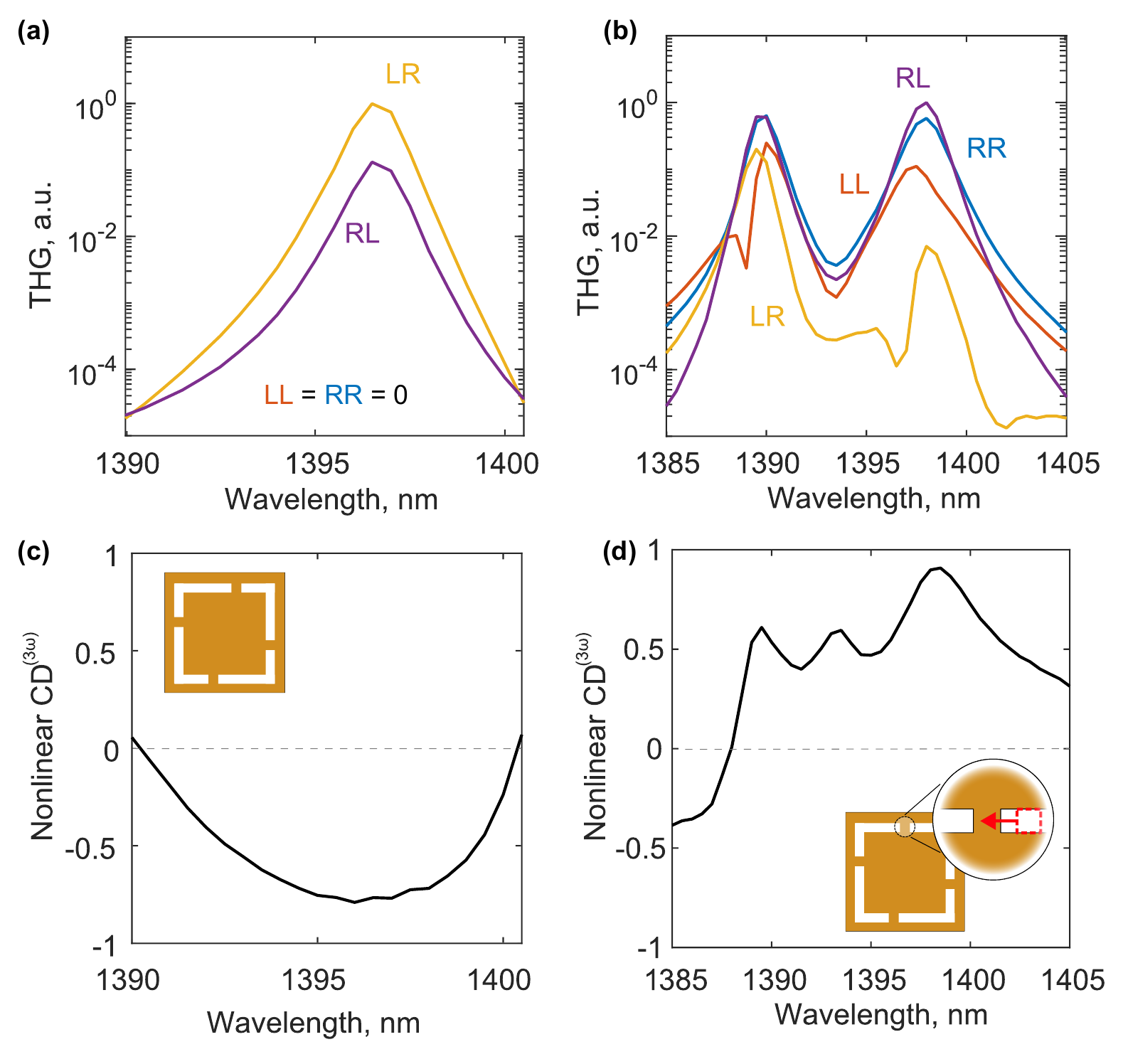}
    \caption{\textbf{Theoretical chiral nonlinear response}. Third-harmonic signals for co-polarized excitation and detection with RCP (blue) and LCP (red), and for cross-polarized signals under LCP (yellow) and RCP (purple) excitation from the unperturbed (a) and perturbed (b) metasurfaces. For unperturbed metasurface co-polarized component LL=RR=0 due to the symmetry. Third-harmonic nonlinear circular dichroism spectra for the unperturbed (c) and perturbed (d) metasurfaces. Insets show schematic images of the corresponding unit cells.}
    \label{fig5}
\end{figure}

To validate the experimentally observed THG results, we simulate the nonlinear response of the metasurfaces. For the modelling, we consider the metasurfaces with out-of-plane mirror symmetry, with full simulation details provided in Supporting Information. As shown in Figure~\ref{fig5}a, the THG from the unperturbed metasurface is strongly enhanced around the structural resonance. In this case, the cross-polarized component LR dominates over the RL component, while both co-polarized components vanish due to the in-plane symmetry. The nonlinear CD is evaluated using Eq.~(\ref{eq2}), with the results presented in Figure~\ref{fig5}c. Near the resonance, CD reaches a maximum magnitude of -0.78, whereas away of the resonance it approaches zero. The nonlinear response of the perturbed metasurface is shown in Figure~\ref{fig5}b. In contrast to the unperturbed case, the THG spectra exhibit two different resonances, consistent with the linear simulations, and the co-polarized components no longer vanish. The corresponding nonlinear CD, shown in Figure~\ref{fig5}d, reveals a change in sign to positive, with a maximum value of 0.91 near the resonance. Overall, the calculated and experimental results exhibit strong agreement. Specifically, for the unperturbed metasurface, both simulation and experiment confirm the dominance cross-polarized component LR and a negative CD at the resonance. For perturbed metasurface, the two peaks THG spectral feature is reproduced in both simulations and experiment, and the sign of CD is consistent between the two. Minor discrepancies, such as experimentally observed finite co-polarized components - absent  in the idealised simulation - are attributed to fabrication imperfection and the influence of the supporting substrate, which effectively breaks the ideal symmetry of the system.

\section{Conclusion}

In conclusion, we have studied both linear and nonlinear properties of free-standing silicon membrane metasurfaces. We have experimentally demonstrated a strong nonlinear chiral response of the metasurface, while such structures exhibit a vanishing linear chiral response due to the out-of-plane symmetry. By introducing structural perturbations that break the $C_4$ symmetry, we enable a significant change in the CD response, tuning it from –0.83 to 0.41. This symmetry-breaking also facilitates the emergence of  co-polarized third-harmonic components, which are otherwise forbidden in the unperturbed metasurface. Our findings suggest a novel approach to engineering nonlinear chiral optical responses in metasurfaces that are effectively achiral in the linear regime, thus paving the way for advanced applications of nonlinear chiral photonics and selective chiral harmonic generation.

\section*{Supporting Information}

Details of sample fabrication, theoretical and experimental methods.

\section*{Acknowledgement}

This work was supported by the Australian Research Council (Grant No. DP210101292) and the International Technology Center Indo-Pacific (ITC IPAC) via Army Research Office (contract FA520923C0023). H.-G.P. acknowledges support from the National Research Foundation of Korea (NRF) grant funded by the Korean Government (MSIT) (2021R1A2C3006781 and RS-2021-NR060087) and the Samsung Science and Technology Foundation (SSTF-BA2401-02). G.L. acknowledges a support from the National Natural Science Foundation of China (12161141010) and the National Key Technologies R$\&$D Program of China (2022YFA1404301). T.T. and T.Y. acknowledge a support from JST CREST (Grant No. JPMJCR1904), Japan. Y.K. thanks Dr. Yongsop Hwang for his initial participation in this project.  I.T. thanks Dr. Kirill Koshelev for valuable advice on the initial design of the metasurface and simulations techniques.

\bibliography{Refs}

\renewcommand\thesection{S\arabic{section}}
\renewcommand\theequation{S\arabic{equation}}
\renewcommand\thefigure{S\arabic{figure}}
\renewcommand\thetable{S\arabic{table}}
\setcounter{figure}{0}
\setcounter{section}{0}

\onecolumngrid
\newpage 
\section*{Supporting Information}

\section{Samples Fabrication}

The membrane metasurfaces were fabricated from a silicon-on-insulator (SOI) wafer consisting of a 370-nm-thick top silicon layer, a buried oxide layer, and a silicon substrate. A positive electron-beam resist (ZEP520A) was uniformly spin-coated onto the top SOI wafer. High-resolution electron-beam lithography was performed to define the designed nanopatterns with sub-100-nm accuracy. The exposed silicon areas were dry-etched using deep silicon reactive ion etching (deep Si RIE). A gas mixture of sulfur hexafluoride (SF$_6$) and a perfluorocarbon (PFC) gas was injected to facilitate anisotropic etching with smooth vertical sidewalls and high aspect ratios. Following dry etching, the remaining ZEP resist was removed by immersion in N-methyl-2-pyrrolidone (NMP) at room temperature overnight, and organic residues were removed using O$_2$ plasma. The freestanding silicon structure was then fabricated by selectively wet-etching the buried oxide layer, immersing the wafer in an NH$_4$F/HF (6:1) buffered oxide etchant (BOE) solution at room temperature.

\section{Numerical Simulations}

All numerical simulations were performed in the Wave Optics module of COMSOL Multiphysics. Linear and nonlinear transmission simulations are simulated in the frequency domain. The metasurface was placed on a semi-infinite air substrate surrounded by a perfectly matched layer mimicking an infinite region in the vertical direction. The simulation area is the unit cell with Floquet periodic boundary conditions which simulate an infinite size of the metasurface in a transverse plane. The dispersion of the refractive index for Si is taken from Ref.~\cite{toftul2024chiral}.
The background field is set manually via custom code using Fresnel equations. The third harmonic generation process is calculated in the undepleted pump approximation using the domain polarization feature~\cite{COMSOL_SHG_example}. The nonlinear polarization current is calculated as $P_{i}^{(3\omega)} = \varepsilon_0 \hat{\chi}^{(3)}_{ijkm} E_{j}^{(\omega)} E_{k}^{(\omega)} E_{m}^{(\omega)}$, where $\hat{\chi}^{(3)}$ tensor has 21 nonzero elements based on the Si point symmetry group $\mathrm{D}_{6 \mathrm{h}}$ (196-th space group)~\cite{boyd2008nonlinear}. We assume that the crystallographic axes are aligned with the metasurface grating direction and incident field direction, i.e. with the base Cartesian unit vectors $(\mathbf{\hat{x}}, \mathbf{\hat{y}}, \mathbf{\hat{z}})$. Once the total fields are calculated for the left- and right-circularly polarized (RCP and LCP) background fields, $\mathbf{E}_{^{\text{R}} _{\text{L}}}$, the  complex transmission amplitude coefficients  are calculated as $t^{(n\omega)}_{^{\text{RR}} _{\text{LL}} } = \braket{\mathbf{\hat{e}}_{\pm}}{\mathbf{E}_{^{\text{R}} _{\text{L}}}^{(n\omega)}} =  \frac{1}{A} \iint\limits_{A} \mathbf{\hat{e}}^{*}_{\pm} \cdot \mathbf{E}_{^{\text{R}} _{\text{L}}}^{(n\omega)} (x,y,z_0) \dd x \dd y$, where $A$ is the area of the $z=z_0$ plane located at the edge of the simulation area in the substrate, and $\mathbf{\hat{e}}_{\pm} = (\mathbf{\hat{x}} \pm \iu \mathbf{\hat{y}} )/\sqrt{2}$ are the unit vector in the circular basis. Integration over surface $A$ averages the output signal over the angles, so it gives only the $0$-th diffraction order. Finally, the transmission coefficients are calculated as $T^{(\omega)}_{^\text{RR} _{\text{LL}}} = \frac{n_{\text{subs}}}{n_{\text{host}}} \abs{t_{^\text{RR} _{\text{LL}}}^{(\omega)}}^2$, and the output nonlinear intensity is $I_{^\text{RR} _{\text{LL}}}^{(3\omega)} \propto \abs{t_{^\text{RR} _{\text{LL}}}^{(3\omega)}}^2$, where the proportionality coefficient is unimportant within the scope of this work. The linear simulation results are shown in Figure~\ref{figS1}.

\begin{figure}[h]
    \centering
    \includegraphics[width=0.65\linewidth]{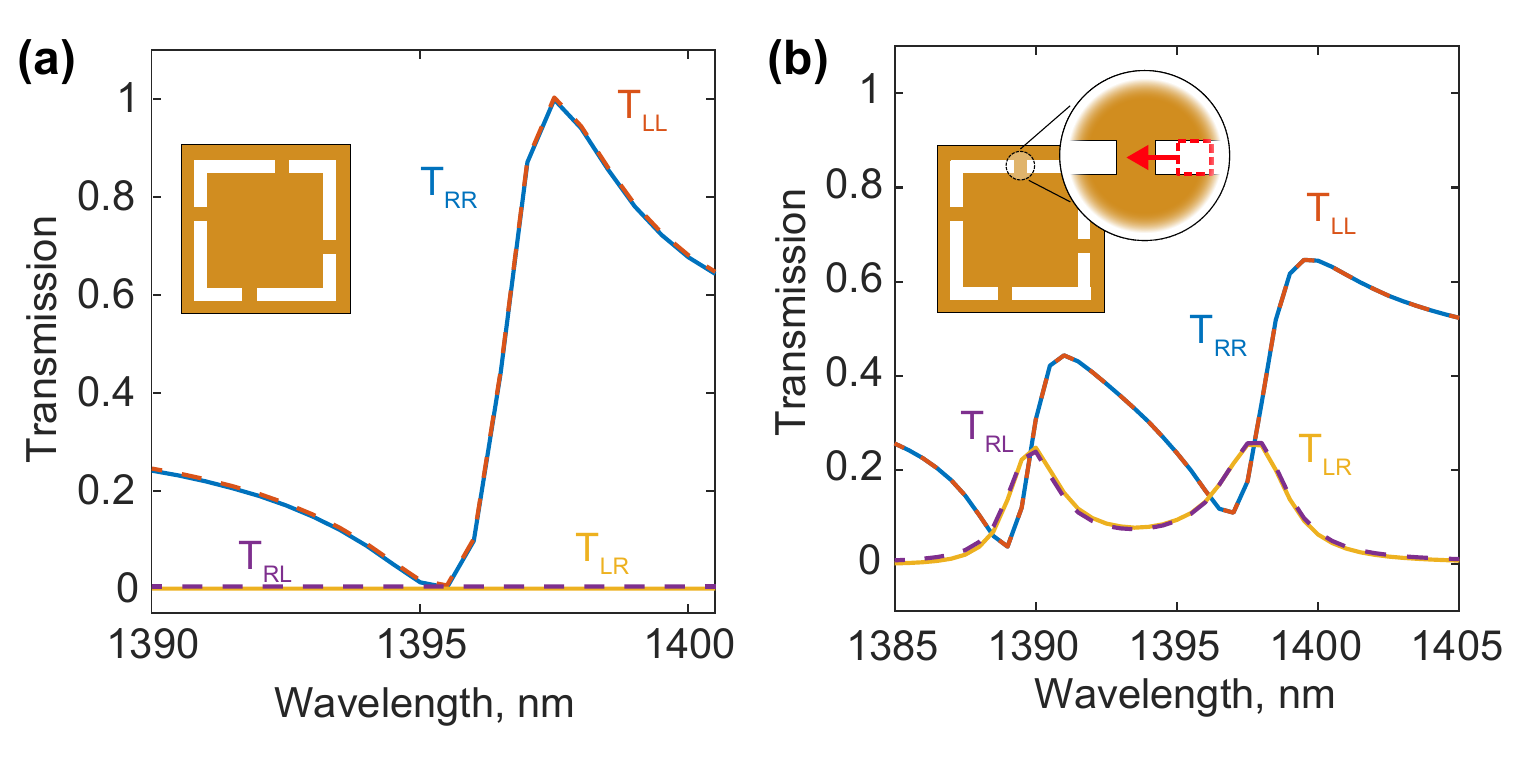}
    \caption{\textbf{Linear optical properties}. Theoretical transmission spectra of the unperturbed (a) and perturbed (b) metasurfaces under left- and right-circularly polarized light.}
    \label{figS1}
\end{figure}

\section{Linear Optical Measurements}

The light source used in the experiment was a QTH10 Quartz Tungsten-Halogen Lamp. The emitted light was first linearly polarized using a wire-grid linear polarizer (Thorlabs WP25M-UB1), and subsequently converted into circularly polarized light by passing through an achromatic quarter-wave plate (Thorlabs AQWP05M-1600). The resulting circularly polarized beam was then focused onto the substrate side of the metasurface membrane using a calcium fluoride (CaF$_2$) lens with a focal length of 50 mm. The transmitted light was collected by a Mitutoyo M Plan Apo 20$\times$ objective lens (numerical aperture NA = 0.4). After collection, the beam was directed through a telescope composed of two achromatic doublet lenses with C-coating, having focal lengths of 75 mm and 150 mm, respectively. An adjustable square-shaped slit (Owis) was positioned at the intermediate focal plane of the telescope to spatially filter the signal, allowing only light transmitted through the metasurface to pass. The filtered light then passed through another set of polarization optics—an achromatic quarter-wave plate and linear polarizer, identical to those used in the input path—to resolve the transmitted light into left- and right-circularly polarized components. Finally, the light was coupled into an optical waveguide using an aspherical lens and directed to a near-infrared spectrometer (NIRQuest, Ocean Optics) for spectral analysis. A schematic of the experimental setup is shown in Figure~\ref{figS2}.

\begin{figure}[h]
    \centering
    \includegraphics[width=0.8\linewidth]{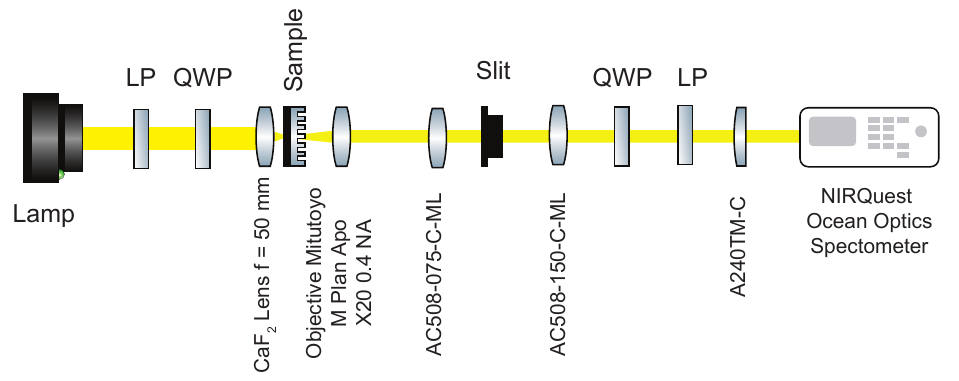}
    \caption{Schematic of the experimental optical setup for linear transmission measurement.}
    \label{figS2}
\end{figure}

\section{Nonlinear Optical Measurements}

In the nonlinear optical measurements, a femtosecond laser-pumped optical parametric oscillator (OPO, Coherent Chameleon; repetition rate: 80 MHz) was used as the excitation source. The pump wavelength is tuned from 1200 nm to 1430 nm. The circular polarization state of the pump wave is controlled by using a linear polarizer (Newport 10GT04) and a quarter-wave plate (Thorlabs AQWP05M-1600). The pump wave is focused onto the metasurface from substrate side by using a 4$\times$ objective lens. The THG signal in the transmission direction is collected by a 10$\times$ objective lens. Subsequently, a quarter-wave plate (Thorlabs AQWP05M-600) and a linear polarizer (Newport 10GT04) are employed to analyze the left- and right- circular polarization components. In addition, a dichroic mirror (Thorlabs DMLP950) and a short-pass color filter (Thorlabs FESH0700) are used to block the pump wave. The THG signal is then measured using an Andor spectrometer (SR-500i) with an EMCCD (electron-multiplying charge-coupled device) detector. In the power dependence measurement, the THG power is obtained by integrating a spectral bandwidth of a 20 nm. Both the THG power and pump power were measured by taking into account of the transmission efficiency of the corresponding optical components. In the wavelength THG responsivity experiment, the pump power for both the unperturbed and perturbed metasurfaces are fixed at 40 mW (before the 4$\times$ objective lens). The peak THG intensity is recorded to calculate the average value and standard deviation for the four circular polarization combinations.  A schematic of the experimental setup is shown in Figure~\ref{figS3}.

\begin{figure}[h]
    \centering
    \includegraphics[width=0.8\linewidth]{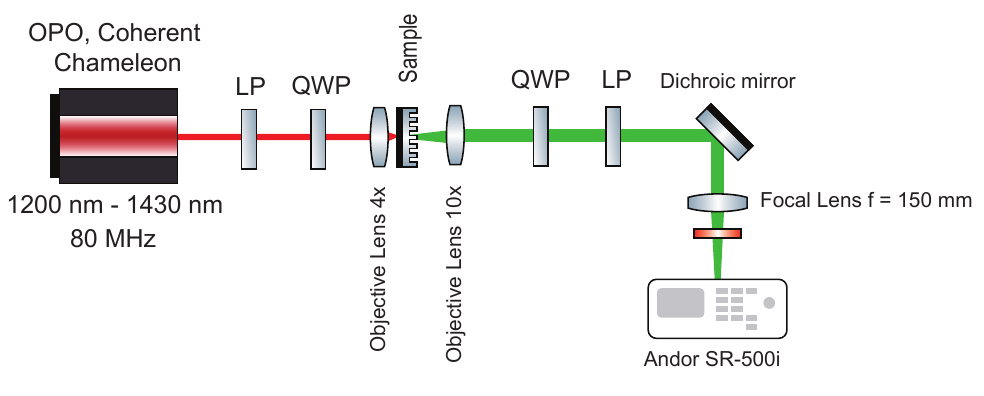}
    \caption{Schematic of the experimental optical setup for nonlinear optical measurement.}
    \label{figS3}
\end{figure}

\end{document}